\documentclass[english,aps,pra,twocolumn,superscriptaddress]{revtex4-1}
\usepackage[latin9]{inputenc}
\usepackage{amsmath}
\usepackage{newtxtext}
\usepackage{newtxmath}
\usepackage{mathrsfs}
\usepackage{tikz}
\usepackage{graphicx}

\makeatletter

\usepackage[colorlinks,citecolor=blue,linkcolor=blue]{hyperref}
\usepackage{amsmath}
\allowdisplaybreaks[2]

\makeatother

\usepackage{babel}

\begin{document}

\title{High-efficiency and noise-immune quantum battery}
\author{Guohui Dong}
\email{dongguohui@sicnu.edu.cn}

\affiliation{College of Physics and Electronic Engineering, Sichuan Normal University,
	Chengdu 610068, China}
	
\author{Mengqi Yu}

\affiliation{College of Physics and Electronic Engineering, Sichuan Normal University,
	Chengdu 610068, China}

\author{Yao Yao}
\email{yaoyao\_mtrc@caep.cn}

\affiliation{Microsystem and Terahertz Research Center, China Academy of Engineering
	Physics, Chengdu 610200, China}

	\begin{abstract}
	Nowadays, quantum batteries (QBs) have been designed
	to outperform their classical counterparts by leveraging quantum
	advantages. For instance, the charging power greatly benefits from the entanglement generation of a collective charging scheme (e.g., the Dicke QB), especially in the ultrastrong
	coupling (USC) regime or even larger. However, apart from the fragility
	of the QB under intrinsic decoherence effects, another critical drawback
	emerges inevitably. Specifically, the non-negligible counter-rotating
	(CR) term in the USC regime would induce coherence in the energy basis
	of QB, thus remarkably degrading the charging efficiency. To tackle
	these challenges, we propose a high-efficiency and noise-immune QB
	boosted by dynamical modulation. It is demonstrated that the time-varying
	modulation can effectively reduce the CR coupling, resulting in a notable improvement in charging
	efficiency. Particularly, for a judicious choice of modulation parameters
	that entirely eliminate the CR interaction, the Dicke QB can be charged
	optimally, resembling the behavior of the Tavis-Cummings QB. In the
	subsequent storage process, beyond the natural robustness to pure
	dephasing noise, our scenario is also highly resilient to the dissipation noise and thus can achieve perfect energy storage by effective bath engineering. While
	feasible with current experimental platforms, our proposal offers
	a solid foundation for the implementation of a powerful QB and may
	drastically promote the development of energy storage and delivery
	techniques in the future.
\end{abstract}
\maketitle

\section{Introduction}\label{sec:intro}

Nowadays, the enormous demands of renewable energy resources, alongside
advancements in mobile technologies such as electric vehicles and
smart phones, have dramatically stimulated the development of energy
storage techniques~\cite{Quach2023,Njema2024}. Considerable attention
has been concentrated on devices or materials with high power and
energy density, such as Li-ion batteries and supercapacitors~\cite{Choi2019,Li2017,Wang2021}.
Recently, inspired by the superior performance of quantum protocols
over their classical counterparts in diverse scenarios~\cite{Gisin2007,Cozzolino2019,Sidhu2021,Steane1998,Ladd2010,Degen2017,Pirandola2018},
combined with the trend of miniaturization over the past century,
quantum batteries (QBs) have been designed to incorporate quantum
principles in the microscopic domain~\cite{Alicki2013}. Motivated
by this interest, significant efforts have been devoted to enhancing
the charging power (the stored energy over charging time) or ergotropy
(the maximal extractable work under unitary dynamics) via uncovering
the role of quantum features (quantum advantage) such as quantum correlations
and coherence in QBs~\cite{Campaioli2017,Ferraro2018,Andolina2019,Francica2020,Shi2022,Campaioli2024,Rinaldi2025}.
For instance, compared to the parallel charging of individual QB cells,
a collective charging scheme can accelerate the charging process and
yield a super-linear scaling of the charging power~\cite{Campaioli2017,Ferraro2018}.
Additionally, the energy transfer speed is also related to the QB-charger
coupling. That is, according to the quantum speed limit theory, the
charging power increases linearly with this coupling strength~\cite{Giovannetti2003,Levitin2009,Deffner2017}.
In this sense, an ultrastrong coupling (USC) collective charging scenario
(or even larger) would remarkably improve the charging power of QBs.

\begin{figure}[t]
	\begin{centering}
		\includegraphics[scale=0.35]{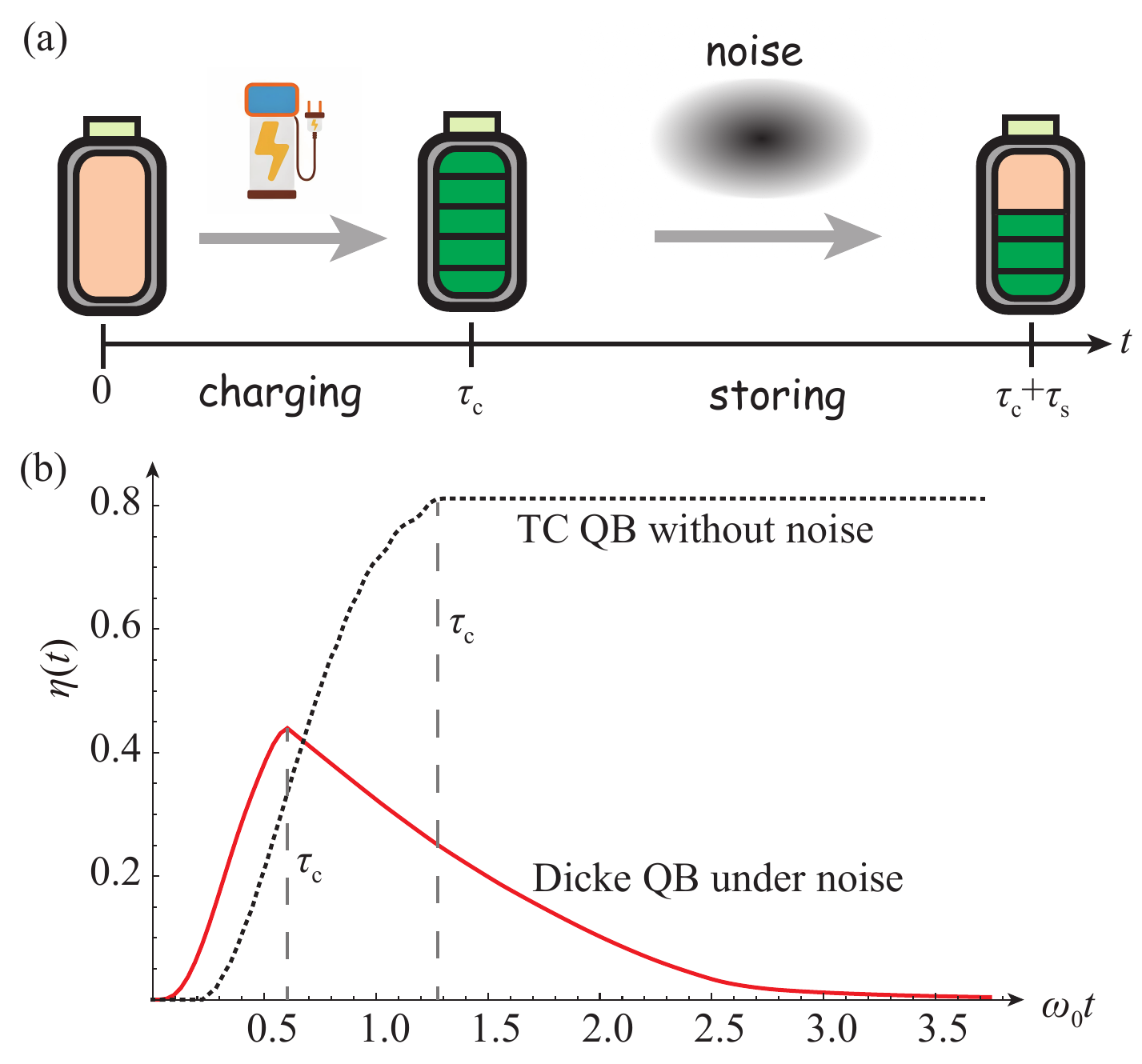}
		\par\end{centering}
	\caption{\label{fig:1}(a) The schematic diagram of the QB in the charging
		and storage process. (b) The efficiency (mean ergotropy) of the QB
		versus time. The red solid line (black dotted line) represents the
		efficiency of the Dicke QB under noise (TC QB without noise). The
		charging time $\tau_{\mathrm{c}}$ is chosen as the one when efficiency
		reaches its first maximum. The initial state of the system is $\left|\varphi(0)\right\rangle =\left|N/2,-N/2\right\rangle \left|N\right\rangle _{c}$.
		Here $g/\omega_{0}=1$, $N=8$, and $\Gamma_{0}/\omega_{0}=0.1$.}
\end{figure}

However, two main drawbacks severely limit the performance of this high-power QB  (see Fig.~\ref{fig:1}). On the one hand, in
the USC regime, the dynamics of the system must be considered beyond
the rotating wave approximation (RWA), which hinders the excitation-conserved
energy transfer between the charger and QB. To be more specific, the
counter-rotating (CR) term, together with the rotating one, would
induce coherence in the energy basis of the battery (see the following
discussions), resulting in a reduction of the efficiency (ergotropy
over the initial energy of the charger). On the other hand, from the
perspective of quantum open systems, the unavoidable interaction between
the QB and its environment may cause the decoherence of the QB, negatively
affecting its energy storage capability~\cite{Farina2019}, which
is also deemed as the self-discharging or aging of the QB~\cite{Kamin2020,Xu2023}.
Generally speaking, two categories of decoherence channels exist for
open systems. The first is pure dephasing, which only diminishes the
coherence of the system in a preferred basis. The second is energy
dissipation, where both the coherence information and energy of the
system flow to the environment~\cite{opensystem2002,Gao2007,Carrega2020}.
To our best knowledge, although plenty of work has been devoted to
mitigating the decoherence effect, most of them focus on only one
noisy channel~\cite{Liu2019,Bai2020,Quach2020}. Therefore, proposals
that can inhibit the CR-induced effect and beat the unwanted impacts
of noise (pure dephasing and energy dissipation) simultaneously are
urgently desired.

In this work, we explore the dynamics of a USC collective-charging
QB under the dynamical modulation (DM) and elucidate the influence
of the DM on the QB explicitly. As illustrated, the time-varying modulation
can effectively reduce the CR coupling while preserving the rotating
term, resulting in a considerable improvement in charging efficiency.
Particularly, for a judicious choice of modulation parameters where
the CR term is fully eliminated, the Dicke QB can be charged optimally,
resembling the behavior of the Tavis-Cummings (TC) QB. In the subsequent
storage process, where energy is saved for future use, beyond its
natural robustness to pure dephasing noise, our scenario is also highly resilient to the dissipation noise and thus can mitigate or even suppress QB's energy loss by effective bath engineering, achieving perfect energy storage.

The remainder of this paper is organized as follows. The model of the DM-assisted collective-charging quantum battery is presented in Sec.~\ref{sec:Model-setup}.  In Sec.~\ref{subsec:Charging}, we explore the energy-transfer process between the QB and charger and demonstrate the promotion of the charging efficiency via the DM. The underlying mechanism of the noise-immune storage under the DM is elucidated in Sec.~\ref{subsec:Noise-Immune-Storage}.   Section \ref{sec:Discussions-and-Conclusions} provides a summary and discussion of our results. The details of derivations are given in Appendixes \ref{subsec:Validtity}-\ref{subsec:master equation}.

\section{\label{sec:Model-setup}Model setup}

In this work, we model the QB as $N$ identical non-interacting two-level
systems (TLSs), which are collectively charged by a cavity mode (referred
to as the charger). Despite the simplicity of a quantum TLS, it has
long been employed as a basic QB cell~\cite{Binder2015,Andolina2018,Bai2020}
given its capability of characterizing the features of extensive models,
for instance, an atom coupled with a photon mode~\cite{scully1997}
or a nitrogen-vacancy (NV) center in diamond~\cite{Acosta2013,Yin2015}.
Here we investigate the QB-charger interaction beyond the RWA, specifically
through the Dicke QB model, to account for a general case of the QB-charger
coupling (strong, ultrastrong, or even larger coupling regimes). Furthermore,
to enrich the controllability of the system, we apply dynamical modulation
signals to both QB and charger, which has been utilized in diverse
systems for various applications, including the decoherence-suppression
schemes~\cite{1999Agarwal}, single-photon transport switches~\cite{Zhou2009},
ultrastrong Jaynes-Cummings model~\cite{Huang2020}, quantum
phase transitions~\cite{Huang2023,Huang2024}, and high-precision quantum estimation
scenarios~\cite{Dong2025}. Moreover, it is worth to emphasize that
the main results of our scenario remains valid irrespective of the QB
size (see the following analysis). Hence our DM-enhanced protocol
is also applicable to the parallel-charging strategy whose features
can be well illustrated by our model with $N=1$~\cite{Ferraro2018,Zhang2019}.

The total Hamiltonian can be cast as $\hat{H}=\hat{H}_{b}+\hat{H}_{c}+\hat{H}_{int}+\hat{H}_{m}$
($\hbar=1$ throughout the work),
\begin{align}
	\hat{H}_{b} & =\omega_{0}\hat{S}_{z},\nonumber \\
	\hat{H}_{c} & =\omega_{c}\hat{c}^{\dagger}\hat{c},\nonumber \\
	\hat{H}_{int} & =\lambda(t)g\hat{S}_{x}(\hat{c}+\hat{c}^{\dagger}),\nonumber \\
	\hat{H}_{m} & =\xi\nu\cos(\nu t)(\hat{S}_{z}+\hat{c}^{\dagger}\hat{c}).\label{eq:HM}
\end{align}
where $\hat{S}_{x(y,z)}=\sum_{i=1}^{N}\hat{\sigma}_{x(y,z),i}/2$
denotes the collective operator of the QB with $i$th Pauli operator
$\hat{\sigma}_{x(y,z),i}$, and $\hat{c}$ ($\hat{c}^{\dagger}$)
is the annihilation (creation) operator of the charger. $\hat{H}_{b}$
and $\hat{H}_{c}$ represent the free Hamiltonians of the QB and charger
with eigen energies $\omega_{0}$ and $\omega_{c}$. $\hat{H}_{int}$
stands for the QB-charger interaction with coupling strength $g$,
which is nonvanishing only in the charging process, i.e., \begin{equation}
	\lambda(t)=\begin{cases}
		1 & 0\leq t\leq\tau_{\mathrm{c}},\\
		0 & t>\tau_{\mathrm{c}}.
	\end{cases}\label{eq:lambda}
\end{equation}
Here $\tau_{\mathrm{c}}$ is the charging time. Without loss of generality,
we assume that the charger is resonant with the QB, e.g., $\omega_{0}=\omega_{c},$
and the Rabi frequency $g$ is real. $\hat{H}_{m}$ is the time-varying
modulation Hamiltonian. Henceforth we regard $\nu$ as the modulation
frequency and $\xi$ as the modulation amplitude.

The ergotropy $\mathcal{E}(t)$ is the maximal extractable work from
the QB via cyclic unitary transformations~\cite{Allahverdyan2004}
\begin{equation}
	\mathcal{E}(t)\equiv\mathrm{Tr}\left[\hat{\rho}_{b}(t)\hat{H}_{b}\right]-\mathrm{Tr}\left[\hat{\tilde{\rho}}_{b}(t)\hat{H}_{b}\right],
\end{equation}
where $\hat{\rho}_{b}(t)\equiv\sum_{k=1}^{d}r_{k}\left|r_{k}\right\rangle \left\langle r_{k}\right|$
is the reduced density matrix of the QB in its eigen basis with $r_{k}\geq r_{k+1}$.
$\hat{\tilde{\rho}}_{b}(t)\equiv\sum_{k=1}^{d}r_{k}\left|\varepsilon_{k}\right\rangle \left\langle \varepsilon_{k}\right|$
is the passive state of $\hat{\rho}_{b}(t)$ in the energy eigenbasis
of $\hat{H}_{b}(t)\equiv\sum_{k=1}^{d}\varepsilon_{k}\left|\varepsilon_{k}\right\rangle \left\langle \varepsilon_{k}\right|$
with $\varepsilon_{k}\leq\varepsilon_{k+1}$. The efficiency $\eta(t)$,
defined as the ergotropy over the initial supply of the charger, characterizes
the ability of the QB of storing energy and then delivering it on
demand, i.e.,
\begin{equation}
	\eta(t)\equiv\mathcal{E}(t)/\mathrm{Tr}\left[\hat{\rho}_{c}(0)\hat{H}_{c}\right],
\end{equation}
where $\hat{\rho}_{c}(0)$ is the initial density matrix of the charger.

\begin{figure*}[t]
	\begin{centering}
		\includegraphics[scale=0.35]{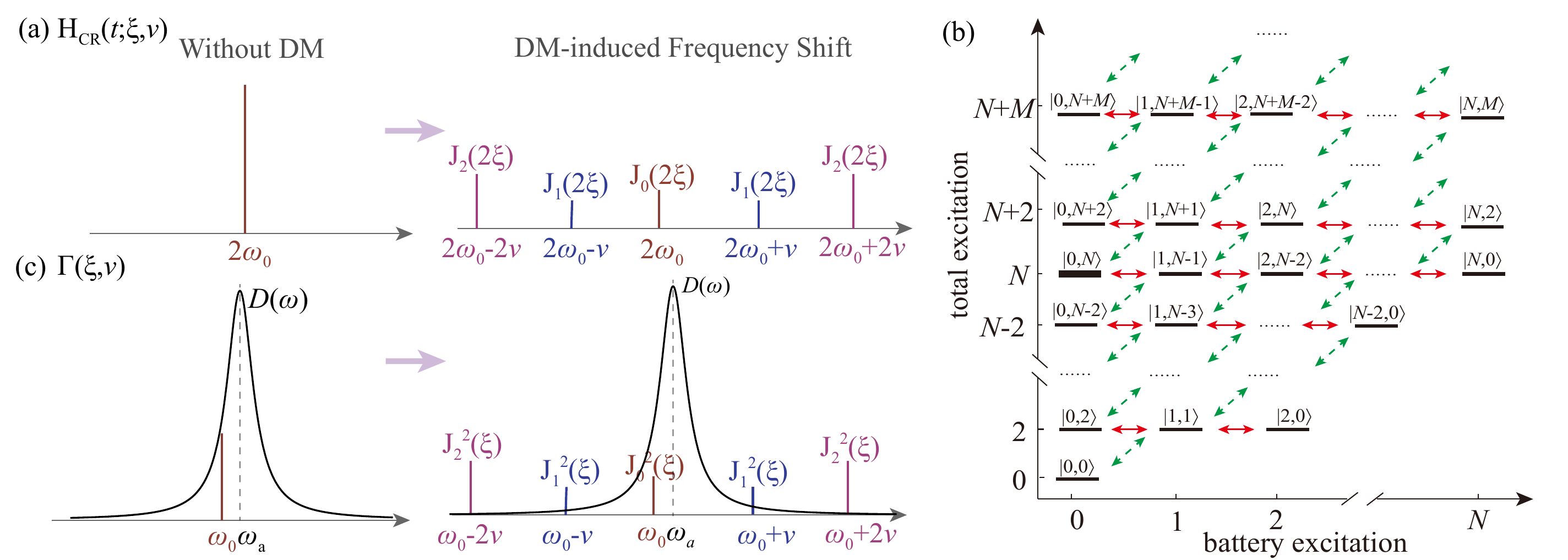}
		\par\end{centering}
	\caption{\label{fig:2} (a) The schematic diagram of the influence of the DM
		on the CR coupling in the charging process. In the large-frequency
		limit, the CR Rabi frequency is reduced from $g$ to $gJ_{0}(2\xi)$.
		(b) The QB-charger transition process. The red solid (green dashed)
		arrows stand for the transitions originating from the rotating (CR)
		interaction. (c) The schematic diagram of the influence of the DM
		on the QB-environment interaction in the storage process. In the large-frequency
		limit, the effective dissipation rate of the QB is shrunk from $\Gamma_{0}\equiv2\pi D(\omega_{0})$
		to $J_{0}^{2}(\xi)\Gamma_{0}$.}
\end{figure*}

In the absence of the DM, the CR term in $\hat{H}_{int}$ and intrinsic
decoherence effect would reduce the battery ergotropy and thus are
detrimental to the performance of the QB. Generally, the charging time
$\tau_{\mathrm{c}}$ in the USC regime is much smaller than the characteristic
lifetime $\tau$ of the system~\cite{FornDiaz2019}, whereas the
storage time $\tau_{\mathrm{s}}$ is far larger than $\tau$. Therefore
we neglect the decoherence of the QB during the charging process. As
demonstrated in Fig.~\ref{fig:1}(b), in the charging process, the
CR term breaks the U(1) symmetry (excitation conservation) and induces
coherence in the QB, causing a lower efficiency (the red solid line)
compared to that of the TC QB, i.e., the collective charging QB under
the RWA (the black dotted line)~\cite{Francica2020,Shi2022}. After
the charging step, the energy or information stored in the QB would
be gradually lost to the environment due to the decoherence effect~\cite{Farina2019,Carrega2020}. Consequently, both strategies (the
CR inhibition and decoherence suppression) would contribute to the
efficiency enhancement (or reservation) of the QB. In the following,
we will elucidate that our DM schemes can achieve these goals and
thus significantly improve the battery behaviour (from the red solid
line to the black dotted line in Fig.~\ref{fig:1}(b)). It is worth
mentioning that the ignorance of the noise effect in the charging
process is more reasonable in our proposal, as the DM can substantially
reduce the QB-environment coupling strength (see Sec.~\ref{subsec:Noise-Immune-Storage}).

\section{DM-boosted Quantum Battery\label{sec:DM-improved-Quanum-Battery}}

In this section, we explore the dynamics of the QB with the DM and
reveal the underlying mechanisms of the improvement in the charging
and storage processes, respectively. In the charging process, the
DM effectively diminishes the coherence in the QB via inhibiting the
impact of the CR coupling and thus increases charging efficiency.
Subsequently, the coherence-suppressed QB would undoubtedly be robust
to the pure dephasing noise even without further modulation. In fact,
for the energy dissipation noise, the modulation can mitigate the
energy loss by reducing the effective QB-environment coupling. More
intriguingly, when the modulation amplitude is chosen as one of the
zeros of a specific Bessel function, the QB is dynamically decoupled
from the environment, yielding perfect storage of the QB energy.

\subsection{High-Efficiency Charging\label{subsec:Charging}}

In the charging process ($0\leq t\leq\tau_{\mathrm{c}}$), this energy
transfer dynamics could be manipulated by the time-varying energy-shift
modulation. In the interaction picture, the interaction Hamiltonian
with respect to $\hat{H}_{0}=\hat{H}_{b}+\hat{H}_{c}+\hat{H}_{m}$
can be cast as 
\begin{align}
	\hat{H}_{I}(t;\xi,\nu) & =\frac{g}{2}\left[\hat{S}_{+}e^{i\omega_{0}t+i\xi\sin(\nu t)}+h.c.\right]\nonumber \\
	& \times\left[\hat{c}e^{-i\omega_{0}t-i\xi\sin(\nu t)}+h.c.\right]\nonumber \\
	& =\frac{g}{2}\left(\hat{S}_{+}\hat{c}+h.c.\right)\nonumber \\
	& +\frac{g}{2}\left[\hat{S}_{-}\hat{c}e^{-2i\omega_{0}t-i2\xi\sin(\nu t)}+h.c.\right]\nonumber \\
	& =\hat{H}_{I,R}+\hat{H}_{I,CR}(t;\xi,\nu),\label{eq:HI}
\end{align}
where the time-independent part $\hat{H}_{I,R}=g\left(\hat{S}_{+}\hat{c}+\hat{c}^{\dagger}\hat{S}_{-}\right)/2$
corresponds to the rotating term while the time-dependent one $\hat{H}_{I,CR}(t;\xi,\nu)$
represents the CR term. It is worth to mention that the Dicke QB converges
to the TC model when the CR interaction in Eq. (\ref{eq:HI}) vanishes.
With the Jacobi-Anger identity $\exp[i\xi\sin(\nu t)]=\sum_{n=-\infty}^{\infty}J_{n}(\xi)\exp(in\nu t)$
where $J_{n}(\xi)$ is the $n$th first kind Bessel function, the
CR term can be rewritten as 
\begin{equation}
	\hat{H}_{I,CR}(t;\xi,\nu)=\frac{g}{2}\sum_{n=-\infty}^{\infty}J_{n}(2\xi)\left[\hat{S}_{-}\hat{c}e^{-i(2\omega_{0}+n\nu)t}+h.c.\right].\label{eq:HI JA identity}
\end{equation}
Eqs. (\ref{eq:HI}) and (\ref{eq:HI JA identity}) indicate that the
modulation shifts the CR interaction to a weighted superposition with
frequencies separated by the modulation frequency (see Fig. \ref{fig:2}(a))
without disturbing the rotating term. Moreover, in the large-frequency
limit ($\nu\gg\omega_{0}$, $g$), the effect of the fast oscillating
term in Eq. (\ref{eq:HI JA identity}) on the system dynamics is negligible.
Hence the interaction Hamiltonian is approximated as (see Appendix
\ref{subsec:Validtity})
\begin{equation}
	\hat{H}_{I}(t;\xi,\nu)\simeq\hat{H}_{I,R}+\frac{g}{2}J_{0}(2\xi)\left[\hat{S}_{-}\hat{c}e^{-2i\omega_{0}t}+h.c.\right].\label{eq:HI JA identity app}
\end{equation}
Obviously, compared to the unmodulated QB, the CR Rabi frequency is
shrunk ($\left|J_{0}(2\xi)\right|\leq1$). That is, the DM effectively
modulates the standard Dicke QB to an anisotropic one with a tunable
CR Rabi frequency.

\begin{figure}[t]
	\begin{centering}
		\includegraphics[scale=0.35]{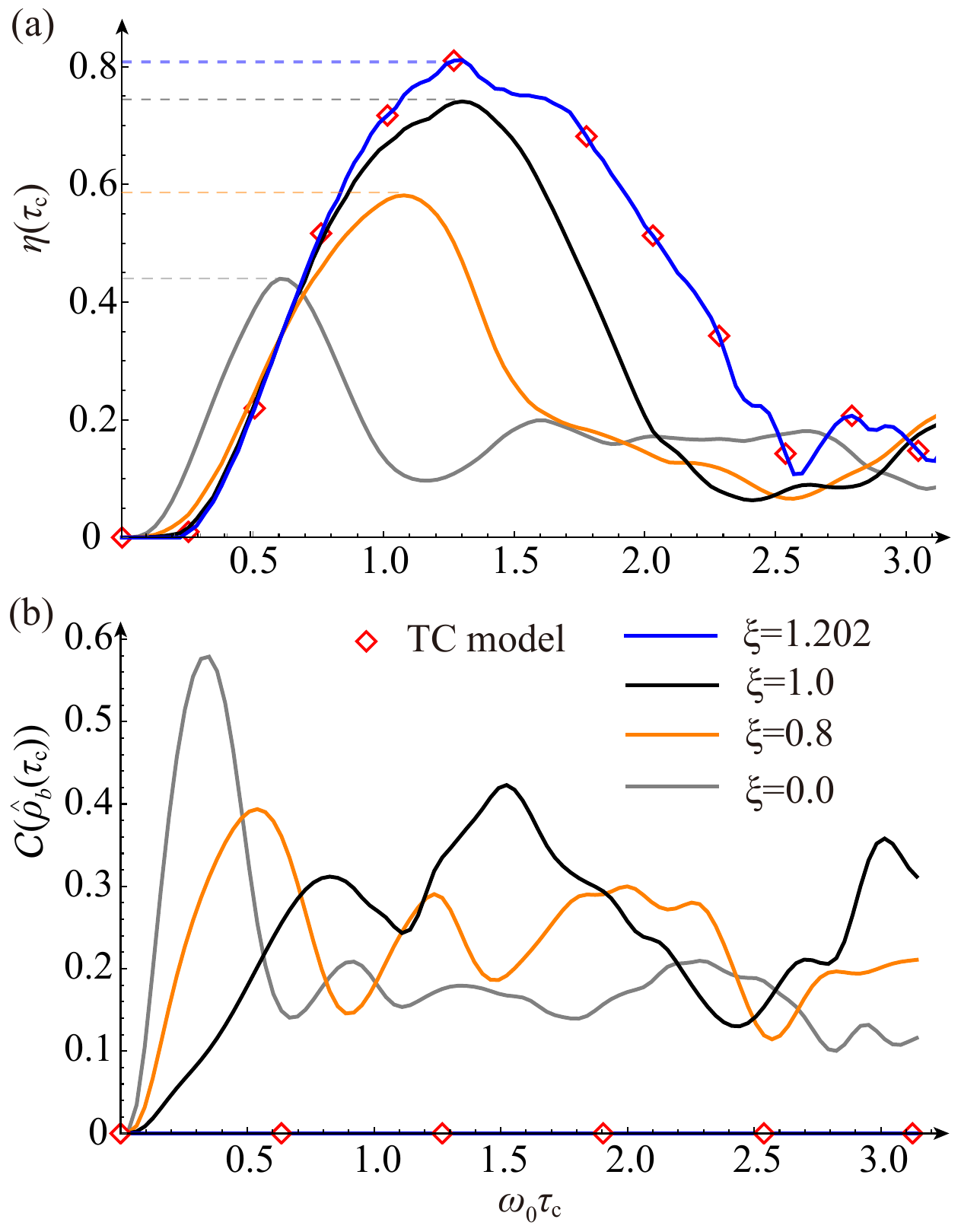}
		\par\end{centering}
	\caption{\label{fig:3}The dynamics of the efficiency (a) and coherence (b)
		of the QB versus charging time. The gray (orange, black and blue)
		solid line represents the numerical result for modulation amplitude
		$\xi=0$ ($0.8$, $1.0$, and $1.202$). The dashed horizontal lines
		mark the maximal efficiencies of the corresponding cases. The red
		diamond denotes the numerical result of the TC QB. Here $N=8$ and
		$g/\omega_{0}=1$.}
\end{figure}	

At the beginning of the charging process, all the QB cells are drained
out of energy (in their ground states) with the charger in its excited
state. Here we assume that the charger is prepared in its Fock state
with the excitation number equaling that of the QB cells, i.e., $\left|\varphi(0)\right\rangle =\left|N/2,-N/2\right\rangle _{b}\left|N\right\rangle _{c}$~\cite{Andolina2019,Lu2021,Rinaldi2025}. Here $\left|S,m\right\rangle _{b}$
$(\left|n\right\rangle _{c})$ is the initial state of the battery
in the Dicke basis (charger in the Fock state) with the total spin
$S$ and magnetic quantum number along the z-axis $m$ (excitation
number $n$). At time $t$, the state of the system evolves

	\begin{figure*}[t]
		\begin{centering}
			\includegraphics[scale=0.35]{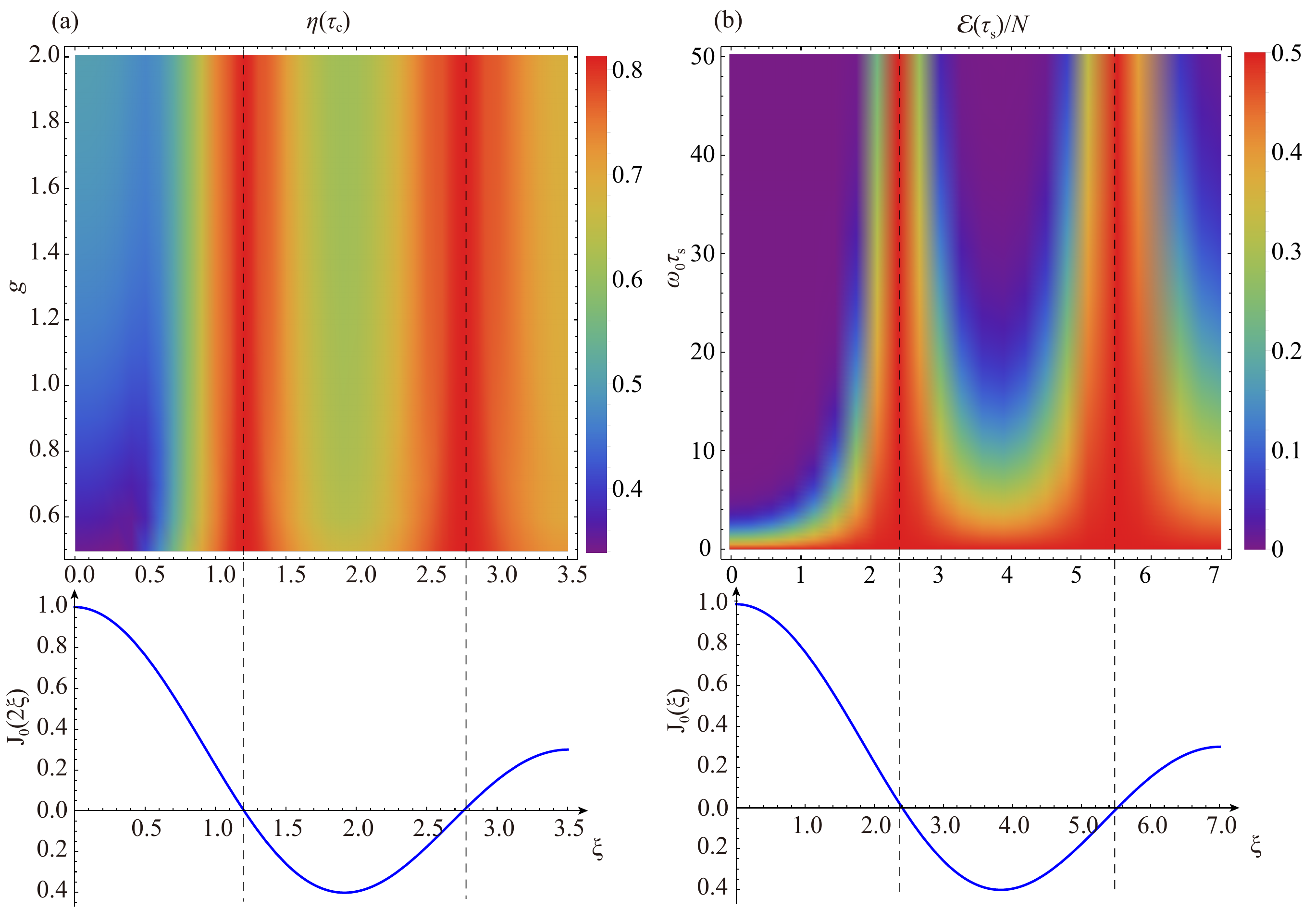}
			\par\end{centering}
		\caption{\label{fig:4}(a) The contour plot of the efficiency of the QB $\eta(\tau_{\mathrm{c}})$
			in the charging process versus the amplitude $\xi$ and Rabi frequency
			$g$. The initial state of the QB-charger system is $\left|N/2,-N/2\right\rangle _{b}\left|N\right\rangle _{c}$.
			(b) The contour plot of the mean ergotropy of the QB $\mathcal{E}(\tau_{\mathrm{s}})/N$
			versus the amplitude $\xi$ and storage time $\tau_{\mathrm{s}}$
			under the energy dissipation noise. The initial state of the QB is
			$\sum^{N}_{m=0}\left|N/2,-N/2+m\right\rangle _{b}/\sqrt{N+1}$. Here
			$N=8$, $\omega_{a}/\omega_{0}=\Omega/\omega_{0}=1$, and $\lambda/\omega_{0}=4$.}
	\end{figure*}

to
\begin{align}
	\left|\varphi(t)\right\rangle  & =\sum_{m=0}^{N}\sum_{n=0}^{\infty}d_{m,n}(t)\left|m,n\right\rangle ,\label{eq:QB charger state t}
\end{align}
where $d_{m,n}(t)$ represents the probability amplitude of $\left|m,n\right\rangle \equiv\left|N/2,-N/2+m\right\rangle _{b}\left|n\right\rangle _{c}$. The dynamical equation of $d_{m,n}(t)$ read
	\begin{align}
		\dot{d}_{m,n}(t) & =-i\frac{g}{2}\sqrt{(\frac{N}{2}-m+1)(\frac{N}{2}+m)}\nonumber \\ & \times \left[\sqrt{n+1}d_{m-1,n+1}(t)+J_{0}(2\xi)\sqrt{n}d_{m-1,n-1}(t)\right]\nonumber \\
		& -i\frac{g}{2}\sqrt{(\frac{N}{2}+m+1)(\frac{N}{2}-m)}\nonumber \\ & \times \left[\sqrt{n}d_{m+1,n-1}(t)+J_{0}(2\xi)\sqrt{n+1}d_{m+1,n+1}(t)\right],\label{eq:equation}
	\end{align}
with the initial condition $d_{0,N}(0)=1$.

The impacts of the rotating and CR couplings on the QB dynamics are
exhibited explicitly in Eq. (\ref{eq:equation}). The rotating interaction
transfers the excitations between the charger and the battery (excitation
conserved). In contrast, the CR coupling connects $\left|m,n\right\rangle $
with $\left|m-1,n-1\right\rangle $ or $\left|m+1,n+1\right\rangle $
where the excitations in the battery and charger are increased by one or
decreased simultaneously. As a result, the coexistence of the rotating
and CR couplings connects states with different excitations and thus induces coherence in the reduced density matrix of
the QB (see Fig.~\ref{fig:2}(b)). Roughly speaking, the coherence
in the energy eigenbasis implies nonzero populations in several states
rather than unity probability in its highest-energy state. Therefore,
although the charging power yields a super-linear scaling in the ultrastrong
Dicke model~\cite{Ferraro2018}, the ergotropy $\mathcal{E}(t)$
and efficiency ($\mathcal{\eta}(t)=\mathcal{E}(t)/N$ in our case)
of the QB would suffer from the nonnegligible CR term.

The DM enhances the charging efficiency of the QB via the effective shrinkage
 of the CR interaction (see Eq. (\ref{eq:HI JA identity app})). As the Bessel function $J_{0}(2\xi)$ varies
with the modulation amplitude $\xi$, the influence of the CR term
on the QB can be tailored by tuning $\xi$. Specially, when $\xi$
is chosen as one of the zeros of $J_{0}(2\xi)$, the CR interaction
disappears. Thus due to
the vanishing of the coherence, the Dicke QB achieves its optimal behaviour, resembling that of the TC model. As demonstrated, the maximal efficiency (Fig. \ref{fig:3}(a))
versus charging time grows from around 0.45 ($\xi=0$, no modulation)
to 0.8 ($\xi=1.202$, zero of $J_{0}(2\xi)$) while the coherence
of the battery (Fig. \ref{fig:3}(b)) is reduced from around 0.6 to
0. Here we employ the relative entropy measure of coherence $\mathcal{C}(\hat{\rho}_{b})\equiv S(\hat{\rho}_{b,dep})-S(\hat{\rho}_{b})$
where $\hat{\rho}_{b,dep}\equiv\sum_{k=1}^{d}\left\langle \varepsilon_{k}\right|\hat{\rho}_{b}\left|\varepsilon_{k}\right\rangle \left|\varepsilon_{k}\right\rangle \left\langle \varepsilon_{k}\right|$
is the dephased state of $\hat{\rho}_{b}$~\cite{Baumgratz2014,Francica2020,Shi2022}.
The effect of the modulation is more clearly in the contour plot Fig.~\ref{fig:4}(a) where the maximal efficiency of the QB $\eta(\tau_{\mathrm{c}})$
is displayed versus the amplitude $\xi$ and Rabi frequency $g$.

	\begin{figure}[t]
	\begin{centering}
		\includegraphics[scale=0.35]{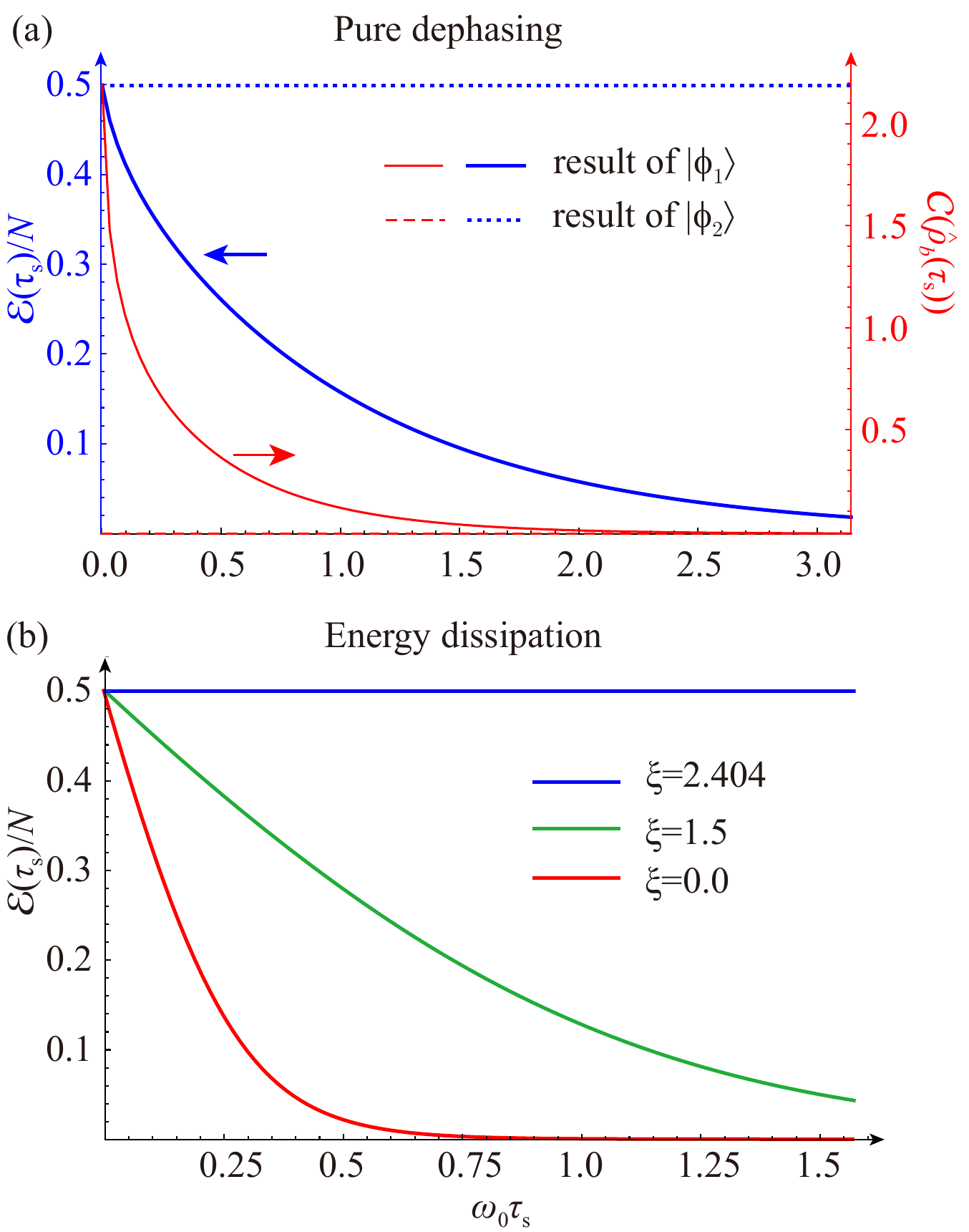}
		\par\end{centering}
	\caption{\label{fig:5}(a) The dynamical evolution of the mean ergotropy (left)
		and coherence (right) of the maximal coherent state $\left|\phi_{1}\right\rangle =\sum^{N}_{m=0}\left|N/2,-N/2+m\right\rangle _{b}/\sqrt{N+1}$
		and incoherent state $\left|\phi_{2}\right\rangle =\left|N/2,0\right\rangle _{b}$
		under the collective dephasing channel. The solid lines (dotted and
		dashed lines) represent the results of the state $\left|\phi_{1}\right\rangle $
		($\left|\phi_{2}\right\rangle $). (b) The dynamical evolution of
		the mean ergotropy of $\left|\phi_{1}\right\rangle $ under the dissipation
		noise. The blue (green and red) lines describes the result of $\xi=2.404$
		($\xi=1.5$ and $\xi=0$). Here $N=8$, $\gamma/\omega_{0}=2$, $\omega_{a}/\omega_{0}=\Omega/\omega_{0}=1$,
		and $\lambda/\omega_{0}=4$.}
\end{figure}

\subsection{Noise-Immune Storage \label{subsec:Noise-Immune-Storage}}

At the end of the charging process, the QB-charger coupling is turned
off, and energy is stored in the QB for future use. During the long-term
storage, the QB-environment interaction can not be neglected, leading
to the decoherence (pure dephasing or dissipation) of the QB~\cite{opensystem2002}.
In the following, we will investigate the energy storage behaviour
of the noisy QB in detail and reveal the robustness of our DM proposal
under two decoherence channels. Namely, beyond the natural immunity
to the dephasing noise, our method can also achieve perfect energy
storage under the energy dissipation channel by effectively washing
out the undesired QB-environment coupling for carefully chosen modulation
parameters.

\subsubsection{Pure Dephasing}

According to the quantum open system theory~\cite{opensystem2002},
the pure dephasing of the QB is induced by its longitudinal interaction
with the environment. Inheriting from the collective nature of our
QB scheme, here we assume that the mean distance between the QB cells
is much smaller than the characteristic length of the environment.
Thus, the Dicke QB loses its phase collectively rather than individually
\cite{Jeske2013,PrasannaVenkatesh2018}. The Hamiltonian can be denoted
as 
\begin{equation}
	\hat{H}_{dep}=\hat{H}_{b}+\hat{H}_{e}+\hat{H}_{1,dep},
\end{equation}
where $\hat{H}_{b}=\omega_{0}\hat{S}_{z}$ is the free Hamiltonian
of the QB (see Eq. (\ref{eq:HM})). The Hamiltonians of the environment
and the QB-environment interaction read
\begin{align}
	\hat{H}_{e} & =\sum_{k}\omega_{k}\hat{a}_{k}^{\dagger}\hat{a}_{k},\nonumber \\
	\hat{H}_{1,dep} & =\sum_{k}(h_{k}\hat{a}_{k}^{\dagger}\hat{S}_{z}+h_{k}^{*}\hat{S}_{z}\hat{a}_{k}),\label{eq:environment}
\end{align}
where $\hat{a}_{k}$ ($\hat{a}_{k}^{\dagger}$) is the annihilation
(creation) operator of the $k$th bosonic bath mode with energies
$\omega_{k}$. $h_{k}$ is the QB-environment coupling strength of
the $k$th mode. The dynamics of the QB under dephasing noise can
be described by the master equation

\begin{equation}
	\dot{\hat{\rho}}_{b}=-i[\omega\hat{S}_{z},\hat{\rho}_{b}]+\gamma\mathscr{L}(\hat{S}_{z}),\label{eq:dephasing ME}
\end{equation}
where $\mathscr{L}(\hat{O})\equiv\hat{O}\hat{\rho}_{b}\hat{O}^{\dagger}-\hat{O}^{\dagger}\hat{O}\hat{\rho}_{b}/2-\hat{\rho}_{b}\hat{O}^{\dagger}\hat{O}/2$
is the Lindblad operator. The dephasing rate $\gamma$ depends on
the QB-environment interaction strength and statistical property of
the environment. Obviously, the dephasing process would degrade the
off-diagonal elements of the QB while maintaining its diagonal entries.
Hence, the ergotropy of the QB with nonzero coherence in the basis
of $\hat{S}_{z}$ would be reduced by the decrease of its coherence~\cite{Francica2020}. On the contrary, the ergotropy of an incoherent
state will be retained in this channel.

Figure~\ref{fig:5}(a) illuminates the above analysis using two states
with the same initial ergotropy ($\left|\phi_{1}\right\rangle =\sum_{m=0}^{N}\left|N/2,-N/2+m\right\rangle _{b}/\sqrt{N+1}$
and $\left|\phi_{2}\right\rangle =\left|N/2,0\right\rangle _{b}$)
but different coherence. On account of the fragility of the coherence
to the dephasing noise, the maximal coherent state $\left|\phi_{1}\right\rangle $
loses its ergotropy (blue thick solid line) rapidly with the decrease
of its coherence (red thin solid line). In the long-time limit, $\left|\phi_{1}\right\rangle $
evolves to the maximally mixed state (identical probability for each
basis $\left|N/2,i\right\rangle _{b}$ $i=-N/2,...,N/2$) with vanishing
ergotropy. As a contrast, the incoherent state $\left|\phi_{2}\right\rangle $
preserves its ergotropy (blue dotted line) all the time due to its
absence of coherence (red dashed line). As we have illustrated, the
coherence of our modulated QB is inhibited in the charging process
(see the blue solid line in Fig. \ref{fig:3}(b)), e.g., $\hat{\rho}_{b}$
is diagonal in the eigenbasis of $\hat{S}_{z}$ at the beginning of
the storage step. In this sense, our dynamically modulated QB is naturally
immune to the pure dephasing noise even without further modulation.

\begin{table}[t]
      \begin{centering}
	\includegraphics[scale=0.35]{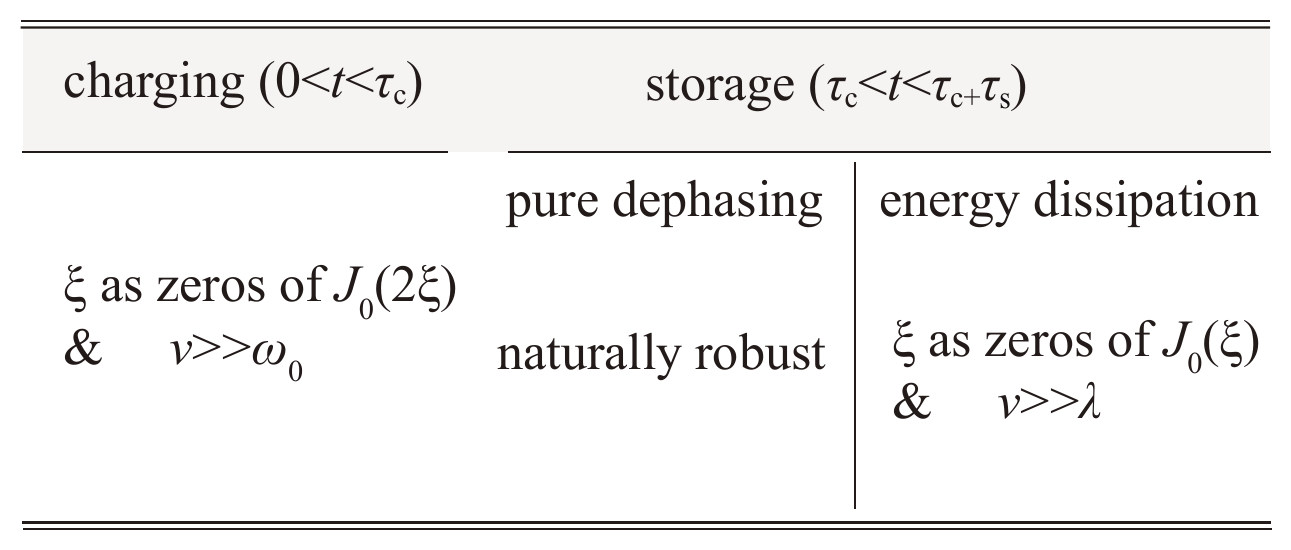}
	\par\end{centering}
\caption{\label{tab:1}The modulation setting for optimal charging and storage
	in our proposal.}
\end{table}

\subsubsection{Energy Dissipation}

As distinguished from the pure dephasing case where only the quantum
coherence is ravaged, the energy dissipation noise would cause the
energy as well as coherence loss of the QB. In this subsection, we try to explicate how our DM scenario
outperforms the unmodulated one under the energy dissipation channel.

Without loss of generality, we model the Hamiltonian of the battery
and environment system with the DM as 
\begin{equation}
	\hat{H}_{dis}=\hat{H}_{b}+\hat{H}_{e}+\hat{H}_{1,dis}+\hat{H}_{m},
\end{equation}
where $\hat{H}_{b}$ and $\hat{H}_{m}$ have been presented in Eq.
(\ref{eq:HM}). The Hamiltonians of the environment and QB-environment
coupling read
\begin{align}
	\hat{H}_{e} & =\sum_{k}\omega_{k}\hat{b}_{k}^{\dagger}\hat{b}_{k},\nonumber \\
	\hat{H}_{1,dis} & =\sum_{k}(f_{k}\hat{b}_{k}^{\dagger}\hat{S}_{-}+f_{k}^{*}\hat{S}_{+}\hat{b}_{k}),
\end{align}
where $\hat{b}_{k}$ ($\hat{b}_{k}^{\dagger}$) is the annihilation
(creation) operator of the $k$th bosonic bath with energies $\omega_{k}$
and coupling strength $f_{k}$.

In the interaction picture with respect to $\hat{H}_{0,dis}=\hat{H}_{b}+\hat{H}_{e}+\hat{H}_{m}$,
the Born-Markov master equation of the QB can be expressed as (see
Appendix \ref{subsec:master equation} for more details)

\begin{equation}
	\dot{\hat{\rho}}_{b}(t)\simeq\Gamma(\xi,\nu)\mathscr{L}(\hat{S}_{-}),\label{eq:dissipation ME}
\end{equation}
where the effective dissipation rate $\Gamma(\xi,\nu)\equiv2\pi\sum_{l=-\infty}^{\infty}J_{l}^{2}(\xi)D(\omega_{l})$
varies with the modulation parameters through the Bessel function
and the shifted spectral density function of the environment $D(\omega)\equiv\sum_{k}\left|f_{k}\right|^{2}\delta(\omega-\omega_{k})$.
$\omega_{l}\equiv\omega_{0}+l\nu$ stands for the $l$th displaced
frequency.

The dynamical modulation effectively engineered the structure of the
spectral density function which then manipulates the decay behaviour
of the QB (see Eq. (\ref{eq:dissipation ME}) and Fig. \ref{fig:2}(c)).
In reality, a reasonable spectral density $D(\omega)$ should vanish
in the large-frequency limit, for instance, the Lorentz spectrum and
the Ohmic family spectral density~\cite{Leggett1987}. Generally
speaking, the width or cutoff frequency of the spectral density characterizes
the inverse of the bath's memory time. Hence, for a modulation frequency
larger than the width or cutoff frequency of $D(\omega)$, the influence
of the environment on the quantum system is distinctly tailored. Namely,
as depicted in Fig. \ref{fig:2}(c), the dissipation rate of the QB
is reduced from $\Gamma_{0}\equiv\Gamma(0,0)=2\pi D(\omega_{0})$
to a series of weighted superposition with frequencies separated by
the modulation frequency $\nu$, i.e., $\Gamma(\xi,\nu)\equiv2\pi\sum_{l=-\infty}^{\infty}J_{l}^{2}(\xi)D(\omega_{l})$.
The weight equals the square of the respective Bessel function. Especially,
for an even larger modulation frequency where the contribution of
the nonzero order terms can be ignored, the effective dissipation
rate approaches that of the unmodulated one with a shrinkage factor
$\Gamma(\xi,\nu)\simeq2\pi J_{0}^{2}(\xi)D(\omega_{0})=J_{0}^{2}(\xi)\Gamma_{0}$.
Apparently, for an appropriate modulation parameter ($\xi$ as one of the
zeros of $J_{0}(\xi)$), the undesired energy and information flow
from the QB to the environment is suppressed entirely, leaving perfect
energy storage.

Figure~\ref{fig:5}(b) plots the dynamics of mean ergotropy of the
QB with dissipation. We figure out the time evolution of the QB (initially
in the maximal coherent state) for different modulation amplitudes.
Here we consider the spectral density as the Lorentz spectrum $D(\omega)=\Omega^{2}\lambda/\left\{ 2\pi\left[(\omega-\omega_{a})^{2}+(\lambda/2){}^{2}\right]\right\} $,
where $\Omega$ denotes the characteristic coupling strength between
the QB and the environment. $\lambda$ ($\omega_{a}$) is the width
(central frequency) of $D(\omega)$. As illustrated in Fig.~\ref{fig:5}(b)
, the modulation effectively reduces the QB-environment coupling strength
and thus mitigates the energy dissipation of the QB (the green and
blue lines higher than the red one). Furthermore, at the zero of $J_{0}(\xi)$,
the ergotropy of the QB keeps undamped due to its effectively decoupling
from the environment (the blue line). Actually, this decoherence suppression
is more evidently in the contour plot of the ergotropy in Fig. \ref{fig:4}(b).

\section{\label{sec:Discussions-and-Conclusions}Discussions and Conclusions}

In our proposal, we have investigated a time-varying field-modulated
quantum battery scheme. Emerging from the DM-induced manipulation
of the QB-charger coupling and bath structure, the dynamics
of the QB can be elaborately engineered on demand. Especially, for
a judicious choice of modulation parameters, our scheme can both enhance
the charging efficiency of the QB in the USC regime beyond the RWA
and simultaneously protect its stored energy from the environmental
noise (see Tab. \ref{tab:1}).

To be more specific, the underlying mechanism can be elucidated as
follows. (i) In the charging process, although a relatively large
Rabi frequency contributes to a quite high charging power, the non-negligible
CR interaction diminishes the charging efficiency by causing coherence
in the reduced density matrix of the QB~\cite{Ferraro2018}. With the modulation employed 
to both the charger and QB, 
the Rabi frequency of the CR interaction shrinks ($gJ_{0}(2\xi)$) while
that of the rotating term remains unperturbed. Notably, at the zero point of the Bessel
function $J_{0}(2\xi)$, the impact of the CR interaction is totally
inhibited, leading to the optimal charging efficiency (as that of
the TC QB). (ii) In the long-term energy storage step after charging,
the effect of the environment (pure dephasing or energy dissipation)
must be taken into account. Beyond the natural robustness of our protocol
to the dephasing noise, which is inherited from the coherence suppression
when charging, our DM scheme can also mitigate the dissipation of the
QB through effectively manipulating the response of the QB to the environment. Similar to the charging process, when $\xi$ is set as one
of the zeros of the Bessel function $J_{0}(\xi)$, the dissipation
of the QB is entirely suppressed, yielding perfect energy storage.

Our DM quantum battery proposal can be implemented in various physical
platforms, for instance, the cavity-QED, circuit-QED, and hybrid quantum
systems~\cite{Kongkhambut2021,Walther2006,Haroche2020,Blais2021,Zou2014}.
For example, in the circuit-QED system, the frequency of a superconducting
qubit as well as transmission-line resonator can be modulated via
changing the flux through a superconducting quantum interference device
(SQUID) loop~\cite{Leek2009,Blais2021,Huang2020}. For an NV center
in diamond, the frequency modulation of the spin states can be achieved
with a time-varying magnetic field. While feasible with current experimental
platforms, our scenario offers a solid foundation for the implementation
of a powerful QB and may drastically promote the development of energy
storage and delivery techniques in the future.
%%%%%%%%%%%%%%%%%%%%%%%%%%%%%%%%%%%%%%%%%%%%%%%%%%%%%%%
%%% Acknowledgements. ??§Ý
%%%%%%%%%%%%%%%%%%%%%%%%%%%%%%%%%%%%%%%%%%%%%%%%%%%%%%%
\begin{acknowledgments}
	The numerical calculation has been performed by using the PYTHON toolbox
	QuTiP~\cite{Johansson2012,Johansson2013}. G.D. and M.Y. are supported by Sichuan Science and Technology Program (Grant No. 2026NSFSC0730) and National Natural Science Foundation of China (Grant No. 12205211). Y.Y.	is supported by National Natural Science Foundation of China (Grant No. 12175204).
\end{acknowledgments}

\section*{DATA AVAILABILITY}
	The data that support the findings of this article are not
	publicly available upon publication because it is not technically
	feasible and/or the cost of preparing, depositing,
	and hosting the data would be prohibitive within the terms
	of this research project. The data are available from the
	authors upon reasonable request.

%%%%%%%%%%%%%%%%%%%%%%%%%%%%%%%%%%%%%%%%%%%%%%%%%%%%%%
%% Appendix sections. ??????, ????
%%%%%%%%%%%%%%%%%%%%%%%%%%%%%%%%%%%%%%%%%%%%%%%%%%%%%%

\appendix

\renewcommand\thesubsection{A\arabic{subsection}}
\setcounter{figure}{0} 
\setcounter{equation}{0} 
\renewcommand\theequation{a\arabic{equation}}
\renewcommand\thefigure{A\arabic{figure}}

\subsection{The validity of Eq. (\ref{eq:HI JA identity app})\label{subsec:Validtity}}

\begin{figure}[t]
	\centering
	\includegraphics[scale=0.35]{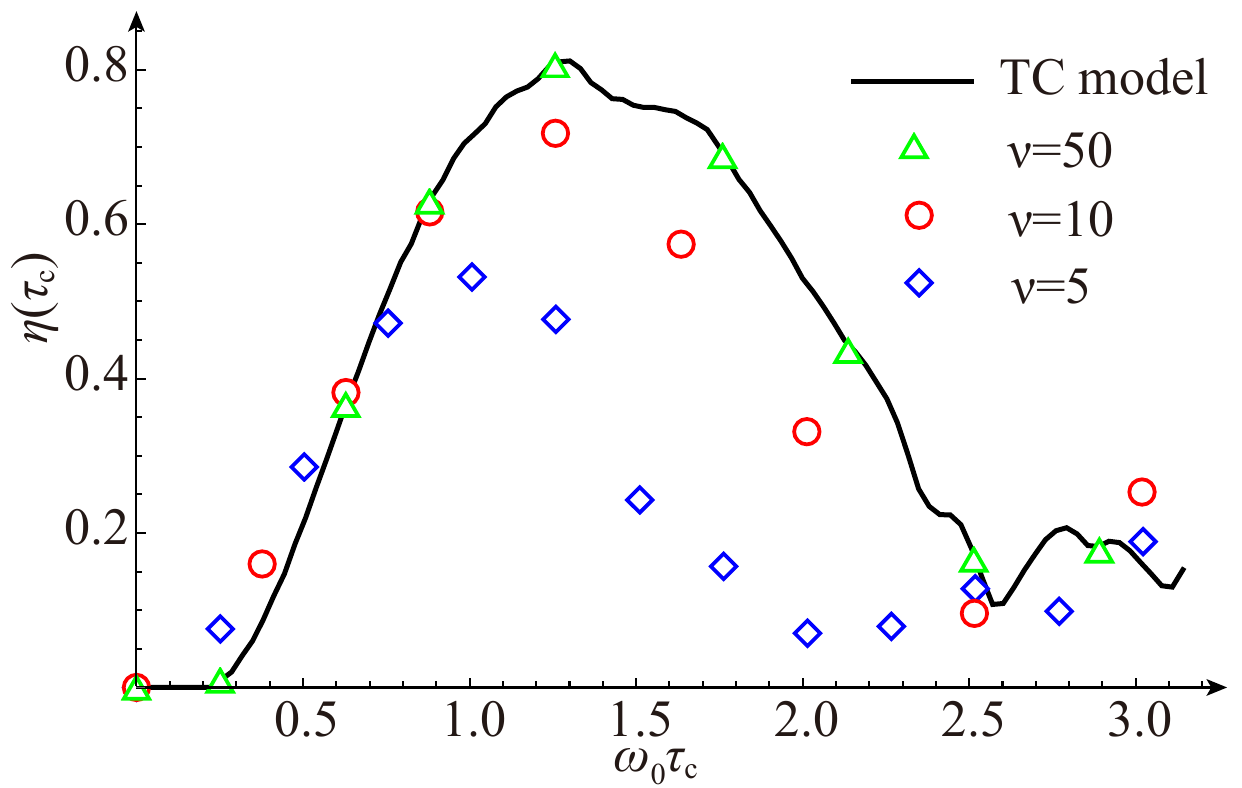}
	\caption{\label{fig:6}The effectiveness of our dynamical modulation proposal.
		The black line represents the efficiency of the TC model while the
		blue diamonds (red circles and green triangles) plot that of the modulated
		Dicke model with $\nu=5$ ($\nu=10$ and $\nu=50$). As illustrated,
		in the large modulation frequency limit, the Hamiltonian of the modulated
		Dicke model can be effectively described by Eq. (\ref{eq:HI JA identity app}).
		Here $N=8$, $g=1$, $\xi=1.202$, and $\omega_{0}=1$.}
\end{figure}

In Fig. \ref{fig:6}, we plot the dynamics of the QB efficiency for
different modulation frequencies (green triangles, red circles, and
blue diamonds). As we have analyzed in Eq. (\ref{eq:HI JA identity app}),
in the large frequency limit, the CR Rabi frequency of the Dicke model
is reduced with a factor $J_{0}(2\xi)$. At the zero of the corresponding
Bessel function, the CR term totally vanishes. In this sense, the
dynamics of the modulated Dicke model resembles that of the TC model
(black line). Fig. \ref{fig:6} demonstrates the coincidence of our
approximated analysis and the numerical calculation results. By setting
$\xi=1.202$ ($J_{0}(2\xi)=0$), we find that the efficiency of the
QB grows with the increase of the modulation frequency. In particular,
for a large enough frequency ($\nu/\omega_{0}=50$), the efficiency
of the modulated Dicke model (green triangles) is the same as that
of the TC model (black line).

\subsection{Master equation\label{subsec:master equation}}

Here we show the derivation of master equation Eq. (\ref{eq:dissipation ME}).
In the interaction picture with respect to $\hat{H}_{0,dis}=\hat{H}_{b}+\hat{H}_{e}+\hat{H}_{m}$,
the interaction Hamiltonian becomes 
\begin{align}
	\hat{H}_{I,dis}(t) & =\sum_{k}f_{k}\left[\hat{b}_{k}^{\dagger}\hat{S}_{-}e^{i(\omega_{k}-\omega)t-i\xi\sin(\nu t)}+h.c.\right]\nonumber \\
	& =\sum_{l=-\infty}^{\infty}J_{l}(\xi)\sum_{k}f_{k}\left[\hat{b}_{k}^{\dagger}\hat{S}_{-}e^{i(\omega_{k}-\omega_{l})t}+h.c.\right]\label{eq:HI B+E interaction picture}
\end{align}
where we have used the Jacobi-Anger identity and defined the notation
$\omega_{l}\equiv\omega+l\nu$.

Up to the second order, the master equation becomes
\begin{equation}
	\dot{\hat{\rho}}_{b}(t)\simeq-\int_{0}^{t}Tr_{E}([\hat{H}_{I,dis}(t),[\hat{H}_{I,dis}(\tau),\hat{\rho}(\tau)]])d\tau,\label{eq:2 order master equ}
\end{equation}
where $\hat{\rho}(t)$ stands for the density matrix of the whole
system. $Tr_{E}(\circ)$ denotes the partial trace over the degrees
of freedom of the environment. Inserting the Hamiltonian in the interaction
picture (Eq. (\ref{eq:HI B+E interaction picture})), we obtain

\begin{widetext}
	
	\begin{eqnarray}
		\dot{\hat{\rho}}_{b}(t)  &\simeq&-\int_{0}^{t}Tr_{E}([\hat{H}_{I,dis}(t),[\hat{H}_{I,dis}(\tau),\hat{\rho}_{b}(t)\otimes\hat{\rho}_{e}]])d\tau \nonumber \\
		&=&\sum_{l,l^{'}=-\infty}^{\infty}J_{l}(\xi)J_{l^{'}}(\xi)\sum_{k,k^{'}}\int_{0}^{t}Tr_{E}\left[(f_{k}\hat{b}_{k}^{\dagger}\hat{S}_{-}e^{-i(\omega_{k}-\omega_{l})t}+h.c.)\hat{\rho}_{b}(t)\otimes\hat{\rho}_{e}(f_{k^{'}}^{*}\hat{S}_{+}\hat{b}_{k^{'}}e^{i(\omega_{k^{'}}-\omega_{l^{'}})\tau}+h.c.)+h.c.\right]d\tau\nonumber \\
		&-&\sum_{l,l^{'}=-\infty}^{\infty}J_{l}(\xi)J_{l^{'}}(\xi)\sum_{k,k^{'}}\int_{0}^{t}Tr_{E}\left[(f_{k}^{*}\hat{S}_{+}\hat{b}_{k}e^{i(\omega_{k}-\omega_{l})t}+h.c.)(f_{k^{'}}\hat{b}_{k^{'}}^{\dagger}\hat{S}_{-}e^{-i(\omega_{k^{'}}-\omega_{l^{'}})\tau}+h.c.)\hat{\rho}_{b}(t)\otimes\hat{\rho}_{e}+h.c.\right]d\tau\nonumber \\
		& =&\sum_{l=-\infty}^{\infty}J_{l}^{2}(\xi)\int_{0}^{\infty}d\omega D(\omega)\int_{0}^{t}d\tau(e^{-i(\omega-\omega_{l})(t-\tau)}+c.c.)\hat{S}_{-}\hat{\rho}_{b}(t)\hat{S}_{+}\nonumber \\
		& -&\sum_{l=-\infty}^{\infty}J_{l}^{2}(\xi)\int_{0}^{\infty}d\omega D(\omega)\int_{0}^{t}d\tau\left[\hat{S}_{+}\hat{S}_{-}\hat{\rho}_{b}(t)e^{i(\omega-\omega_{l})(t-\tau)}+\hat{\rho}_{b}(t)\hat{S}_{+}\hat{S}_{-}e^{-i(\omega-\omega_{l})(t-\tau)}\right].\label{eq:ME1}
	\end{eqnarray}
\end{widetext}

In the above derivation, beyond the typical approximations
utilized for the Markovian master equation in quantum open system
theory, i.e., the Born and Markovian approximations, we have also
neglected the contribution of the fast oscillating terms, e.g., the
off-diagonal terms for the summation over $l$ and $l^{'}$ in the
large-frequency limit. In fact, the effectiveness of this approximation
has been demonstrated in Appendix \ref{subsec:Validtity}.

In the long-time limit, Eq. (\ref{eq:ME1}) can be simplified as

\begin{align}
	\dot{\hat{\rho}}_{b}(t) & \simeq2\pi\sum_{l=-\infty}^{\infty}J_{l}^{2}(\xi)D(\omega_{l})\hat{S}_{-}\hat{\rho}_{b}(t)\hat{S}_{+}\nonumber \\
	& -\pi\sum_{l=-\infty}^{\infty}J_{l}^{2}(\xi)D(\omega_{l})\left[\hat{S}_{+}\hat{S}_{-}\hat{\rho}_{b}(t)+\hat{\rho}_{b}(t)\hat{S}_{+}\hat{S}_{-}\right],\label{eq:ME2}
\end{align}
which is Eq. (\ref{eq:dissipation ME}) in the main text. Here we
have ignored the Lamb-shift Hamiltonian~\cite{opensystem2002}.

\end{document}